# Restoring highly corrupted images by impulse noise using radial basis functions interpolation

**Fariborz Taherkhani, Mansour Jamzad**

*Department of Computer Engineering, Sharif University of Technology, Tehran, Iran*

Fariborz@uwm.edu, Jamzad@sharif.edu

**Abstract**

Preserving details in restoring images highly corrupted by impulsive salt and pepper noise remains a challenging problem. We proposed an algorithm based on radial basis functions (RBF) interpolation which estimates the intensities of corrupted pixels by their neighbors. In this algorithm, first intensity values of noisy pixels in the corrupted image are estimated using RBFs. Next, the image is smoothed. The proposed algorithm can effectively remove the highly dense impulsive salt and pepper noise. Experimental results show the superiority of the proposed algorithm in comparison to the recent similar methods, both in noise suppression and detail preservation. Extensive simulations show better results in measure of peak signal-to-noise ratio (PSNR) and structural similarity index (SSIM), especially when the image is corrupted by very highly dense impulse noise.

## 1. Introduction

Noise removal is a common pre-processing step to improve the visual property of the recovered signal, in this case the image. During image acquisition or transmission, digital images could be degraded by impulse noise. Two common types of impulses are the salt and pepper noise and random-valued noise[1, 2]. Impulsive salt and pepper noise is a special kind of noise that takes place for a short duration with high energy because of camera sensors or transmission in noisy channels. Many algorithms have been suggested to remove this type of noise with high quality in terms of PSNR and SSIM. There are many algorithms to remove these noises while preserving image details. It is known that if the noise is not additive, linear filtering fails, so most algorithms use non-linear approaches to get better results. Median filter (MED) and its modified versions are the most popular methods used in literature because of their de-noising power and computational efficiency [3, 4]. One of the modified versions of MED is Adaptive Median Filter (AMED) [5], in which the window size changes to find a non-noisy pixel as median, but this type of process is time-consuming and error propagating. In addition, it ends up losing the real edges for highly corrupted images. To overcome these problems, symmetric and asymmetric trimmed median filters, such as method [6] and modified decision based unsymmetrical trimmed median filter (MDBUTMF) [7], have been developed.

Impulse detectors play an important role in noise suppression. False detection may result in blurring edges; in addition, such miss detection may leave behind some noisy pixels on the filtered results. In [1], noise candidates are detected by an adaptive median filter and then the image is restored by using a specialized regularization approach that just applies on the selected noisy pixels. Method [8] has introduced a gradual learning method to improve noise detection accuracy and then reduces the noise with a partial differential equation (PDE) in-painting method. In order to improve noise detection accuracy, many other algorithms are proposed, such as the effective decision-based algorithm (EDBA) [9] and its modified version, improved decision based algorithm (IDBA) [10]. These methods identify noise candidates by comparing the center pixel with the minimum and maximum intensity value in a $(3 \times 3)$ window. In [11], the noisy pixels are detected by increasing the ordered difference between the central pixel and its neighbors. In [12] salt and pepper pixels are determined by the process of gradual learning. The fuzzy impulse noise detection and reduction method (FIDRM) filter utilizes the fuzzy gradient value to detect if a pixel is corrupted or not[13]. Experimental results show that these methods [8,11,12,13] apply impulse detectors which have higher computational complexity and lower efficiency in terms of PSNR and SSIM in comparison to some recent efficient approaches such as efficient weighted-average filtering (EWA) [14]. Some of the recent research in this regard apply two steps algorithm for impulse noise removal; the first step is impulse detector and the second step is filtering.

EWA [14] uses min-max impulse detection and then it applies nearest neighboring interpolation and weighted-average filter to restore corrupted intensity values of noisy pixels.

Analysis prior (AP) [15] is one of the non-linear approaches that uses split-Bregman technique to remove impulse noise in corrupted images. The method formulates the noise removal problem in terms of minimizing a $L_p$- regularized and $L_q$- norm data mismatch and then uses split-Bregman techniques to estimate intensity value of noisy pixels.

Progressive switching median (PSM) [16] enhances the quality of the corrupted images gradually and iteratively in which both impulse detector and noise filter are applied progressively and iteratively. This method uses switching schema impulse noise detector and then filters out only a portion of the noisy pixels and leaves behind some noisy pixels on the filtered results; this method applies this step progressively through several iterations to filter out whole image. The important advantage of this approach is that some noisy pixels located in the middle of a large noisy region can be properly detected and filtered.

Iterative adaptive fuzzy filter using alpha-trimmed mean ( AIFATM) [17] uses a fuzzy approach for impulse detection and restoration of corrupted pixels, it applies a weighted mean filtering operation on nearby uncorrupted pixels. Modified switching bilateral filter (MSBF) [18] uses an interval on gray level intensity values for impulse detector. This method uses properties of bilateral filter to preserve detail and edge information in the corrupted image. It uses global trimmed mean with modified switching bilateral filter to restore corrupted pixels. In this method, edge detector is used to enhance the fine details due to edge preserving properties of the bilateral filter.

Spatially adaptive-iteratively reweighted norm (SA-IRN) [19] is a non-linear approach which is not required to know about the statistic of the noise in a corrupted image. In this method, after min-max impulse detection it performs an iterative minimization cost function based on iteratively reweighted norm (IRN) algorithm to estimate the corrupted noisy pixels.

Noise adaptive fuzzy switching median (NAFSM) [20] does not require fine tuning or a learning parameter to get better results; this approach uses the histogram of the corrupted image to detect noisy pixels and then uses fuzzy switching median filter to restore corrupted pixels.

In the proposed method, we estimate intensity value of noisy pixel in a corrupted image by impulsive salt and pepper noise using radial basis functions (RBFs). These functions have been found to be widely successful for interpolation of scattered data. Generally, the correlation of the pixels in a small region of an image is high and the difference of their intensity values is low. Therefore, in this case with negligibility, we can interpolate the intensity values of pixels in a small region with continuous functions such as RBFs. In this method, we fit a continuous model in a window. The center of window is positioned on a noisy pixel such that it has the least possible size while containing at least one non-noisy pixel.

After interpolating the intensity values of noisy pixels, we smooth the image with weighted coefficients that are determined by Euclidean distance between the noisy and non-noisy pixels. We perform smoothing because it is probable that artificial edges were created in the interpolation step and also it is possible that we have faced over fitting problem [21] in estimation of noisy pixels intensity values. We tested the proposed algorithm on four standard images that are corrupted by salt-and-pepper noise with a wide range of noise levels varied from 10% to 95%. The results are compared with other well-known filters, including the MED, EDBA [9], EWA [14], PSM [16], AIFATM [17] , AP [15], MDBUTMF [7] and NAFSM [20]. The experimental results have shown that the proposed algorithm performs effectively both in noise suppression and edge preservation. Although the proposed filter is simple and does not require tuning any parameter to improve the results, it can preserve image edges surprisingly well. The results show that in the measure of peak signal to noise ratio (PSNR) and structural similarity index (SSIM) [28]; this method has better performance rather than other algorithms, especially in high ratio of corrupted image.

The rest of paper is organized as follows. In Section 2 we explain the proposed algorithm. In Section 3, the experimental results are discussed. The conclusion is presented in Section 4.

## 2. Proposed algorithm

In the classical salt and pepper noise, the statistic of the image is formulated as equation (1). In this model *f(x)* represents corrupted pixel located at *x*, *(p+q)* is level of noise in image and *R(x)* denotes uncorrupted pixel located at *x*. Since we use 8 bits gray scale images, $L_{min} = 0$ and $L_{max} = 255$.

$$f(x) = \begin{cases} L_{min} & \text{with probability of } q \\ L_{max} & \text{with probability of } p \\ R(x) & \text{with probability of } 1 - (p + q) \end{cases} \quad (1)$$

We estimate the intensity value of noisy pixels by their non-noisy neighbors. We consider the fact that, generally, in an image each small region has a similar property within itself and there is little alternation of information in its adjacent neighborhoods. Therefore, the correlation of pixels in a small region is high and the difference of intensity values of its pixels is low [22]. Therefore, with negligibility, for intensity values of these pixels, we can have continuous functions of their locations and we can fit a continuous model for this region. To fit a continuous model, we use Radial Basis Functions (RBFs). We designed an equation that estimates the intensity value of a noisy pixel based on its position and uses it as the input for our intensity estimator function. Therefore, we define a function *F(x)* that represents the intensity estimation of noisy pixel at position *x*. To define *F(x)*, we use a linear combination of (RBFs) [23, 24]. The proposed algorithm is described in the following three steps.

### Step 1. Designing the estimation function

In this step, we estimate the intensity value of the noisy pixels using RBFs and produce an initial version of the restored image. A radial function is a radially symmetric function around a point $X_c$ which is called the function's center. RBF is a real-valued function whose value depends only on the distance from the center. One of the characteristics of RBFs is that their responses change monotonically when the distance from central point changes. Therefore, this characteristic is suitable to fit a continuous model for regions in which there is little alternation of information. The parameters of this model are the distance scale and the precise shape of the radial function. With assumption of linear model, all parameters are fixed. A specific kernel Q in an estimation equation such as (2), defined in the form of Q: $R^d \times R^d \rightarrow R$, is considered as a radial function if $Q(x, x_j)$ is a function of Euclidean distance between $x$ and $x_j$.

In this case $x$ is the input of kernel which is the location of noisy pixel and $x_j$ is the location of non-noisy pixel which is the central point used for setting the radial function on it.

We approximate the intensity values of noisy pixels as follows: given non-noisy pixels $x_1$, $x_2,...,x_n$ in $R^d$, construct a function $F: R^d \rightarrow R$ such that $F(x_i) = y_i$ for $i=1,2,...n$. In our case for image restoration, $d=2$ and $y_i$ is the intensity value of pixel located at $x_i$. $F(x)$, the intensity value estimator of noisy pixel located at $x$ is constructed using a linear combination of RBFs. The center point of a RBF is aligned on each non-noisy pixel. If we have $x_1, x_2,...,x_n$ non-noisy pixels in a small region, we can construct $F$ with a linear combination of their corresponding $n$ RBFs as follows:

$$F(x) = \sum_{j=1}^{n} c_j Q\left(\left\| x - x_j \right\|_2\right) \quad (2)$$

where $//x - x_j//_2$ is the Euclidean distance between $x$ and $x_j$, the constant coefficients $c_j$ are obtained by ensuring that the estimator function precisely matches the intensity values of non-noisy pixels to their locations. This is obtained by enforcing $F(x_i) = y_i$ which creates the following linear equations:

$$QC = Y; \quad (3)$$

Where

$$\mathbf{Q} = \begin{pmatrix} Q(\|x_1 - x_1\|_2) & Q(\|x_1 - x_2\|_2) & \cdots & Q(\|x_1 - x_n\|_2) \\ Q(\|x_2 - x_1\|_2) & Q(\|x_2 - x_2\|_2) & \cdots & Q(\|x_2 - x_n\|_2) \\ \vdots & \vdots & & \vdots \\ Q(\|x_n - x_1\|_2) & Q(\|x_n - x_2\|_2) & \cdots & Q(\|x_n - x_n\|_2) \end{pmatrix} \quad (4)$$

$$C = [c_1, c_2, ..., c_n]^T \quad (5)$$
$$Y = [y_1, y_2, ..., y_n]^T \quad (6)$$

We find $\{c_1, c_2, ..., c_n\}$ coefficients by solving the linear equation (3). To have a unique solution for equation (3), the interpolation matrix $Q$ should be symmetric positive definite. All eigenvalues of a positive definite matrix are positive. Therefore, a positive definite matrix is invertible [25]. In the following steps, we describe how to solve our problem using these properties.

**Step 2. Interpolation**

We scan the image, if a pixel is noisy (i.e. its intensity value is either 0 or 255) then we define a 3×3 window centered on it. If all pixels in this window are noisy, we increase the window size until we find at least one pixel which is not noisy. All non-noisy pixels in a window are used to estimate the intensity values of noisy pixels in the same window.

One of the reasons that we stop increasing window size after finding at least one non-noisy pixel, is that if we don't do so, the region that we want to fit a continues model becomes larger. Therefore, in this larger region, there would be a greater difference between pixels intensities. It means that, the pixels inside this region would have less correlation in their intensities. This contradicts our assumption that a small region has homogeneity in intensity. Thus, in a large window, a continuous function can't properly estimate the intensity values of noisy pixels.

Limiting window size also helps us to preserve edge detail information. In this step, we record the window size determined for every noisy pixel which is then used in smoothing step (third step). Now we can formulate the estimation process. We choose every non-noisy-pixel in the defined window as a center data to construct the interpolation matrix Q. Suppose $N = \{x_1, x_2, ..., x_n\}$ are positions of these $n$ non-noisy pixels in the window and $I = \{y_1, y_2, ..., y_n\}$ are their intensity values. To construct the interpolation matrix, we should choose a radial basis function. Type of this radial function influences on the performance of the interpolation. Therefore, we experimentally selected inverse quadric kernel as defined in equation (7).

$$Q(r_{i,j}) = 1/(1 + (\varepsilon\, r_{i,j})^2) \tag{7}$$

$$r_{i,j} = \| x_i - x_j \|_2 \qquad \forall\, x_i, x_j \in N \tag{8}$$

where $\varepsilon$ is a free shape parameter that plays an important role for the accuracy of approximation [24]. We initiate this parameter as suggested in [26, 27]:

$$\varepsilon = \frac{0.8\sqrt{n}}{w} \tag{9}$$

where $n$ is the number of members in a set of non-noisy pixels $N$, and $w$ is window size, and $r$ is Euclidean distance between two points $x_i$ and $x_j$ in set $N$. We normalize the data to get better performance as well as to be able to implement the proposed method in the computer because the elements of the matrix $Q^{-1}$ in (11) are very huge numbers; therefore, we change the intensity values in set $I$ to smaller numbers using an exponential form same as (10):

$$Y_i = \exp(-y_i) \tag{10}$$

Now, we can construct the interpolation matrix Q which is defined in (4). Based on (3), our unknown parameters are $C = \{c_1, c_2, ..., c_n\}$ coefficients. We can find $C$ by the following equation which has a unique solution.

$$C = Q^{(-1)} Y \tag{11}$$

Now we can estimate the intensity value of noisy pixel by equation (2). Since we have changed the intensity values in (10), we should change $F(x)$ to $f_i$ as follows:

$$f_i = -\ln(F(x)) \tag{12}$$

**Step 3. Smoothing**

Because of residual error resulted from interpolation step as well as the possibility of over fitting problem which is a common concern in scattered data interpolation [5], in this case estimation of the noisy pixels intensity value, it is highly probable that artificial edges are created in step 2. To remove artificial edges and improve preserving edges and image details, we perform this step only on the pixels determined to be noisy in the step 2. We smooth the intensity values of noisy pixels with weighted coefficients related to the Euclidean distance ($r_{i,j}$ in eq.14) between the noisy and non-noisy pixels in the window defined in step 2. A noisy pixel is smoothed as follows:

$$f_i = \frac{\sum_{j \in \text{pixels in the window}} f_j \times exp(-\alpha r_{i,j})}{\sum_{j \in \text{pixels in the window}} exp(-\alpha r_{i,j})} \quad (13)$$

where $f_i$ is the smoothed value of noisy pixel, $f_j$ represents the intensity of a pixel in defined window, and $r_{i,j}$ is the Euclidean distance between the noisy pixel $i$ and any other pixel $j$ in the defined window. Parameter α regulates impact of the distance in equation (13). In the proposed algorithm, we experimentally set α=1. Based on (13), as much as the distance between noisy pixel and other pixel in the window increases; the corresponding weight of that pixel in equation (13) decreases as well.

By the end of smoothing step, we might have some outliers remained inside the defined window. An outlier data is defined as follow: Let μ and $\sigma$ be the mean and variance of pixels' intensity values in the window. If the intensity value of a pixel in the defined window is greater than $\mu + 2\sigma$ or smaller than $\mu - 2\sigma$, then that pixel is considered as an outlier.

To reduce the degradation effect of outliers, we replace every outlier data to the median value of pixels in the defined window. However, in the case that median remains to be an outlier itself, then there will be no change on the outlier pixel. In practice, this replacement happens rarely, because intensity values in the window are smoothed by Gaussian kernels so the probability of the intensity values in window that fall into the outlier region (i.e. $f_i > \mu + 2\sigma$ or $f_i < \mu - 2\sigma$) is low and probability of median being an outlier is low as well. We perform this replacement to reduce artifact of over fitted data as much as possible.

**3. Experimental results**

In the following, we report the performance of the proposed algorithm and compare it with four other well-known filters including the MDBUTMF[6], EDBA[9], EWA[14] and AIFATM[17]. In some of these filters, parameters are set based on the suggestions of authors.

We evaluate quantitatively the performance of all filters with PSNR and structural similarity index measure (SSIM) [28].

$$MSE = \frac{\sum_i \sum_j (X(i,j) - Y(i,j))^2}{M \times N} \tag{14}$$

$$PSNR = 10 \log \frac{(255)^2}{MSE} \tag{15}$$

$$SSIM = \frac{(2\mu_x \mu_y + c_1)(2\partial_{xy} + c_2)}{(\mu_x^2 + \mu_y^2 + c_1)(\partial_x^2 + \partial_y^2 + c_2)} \tag{16}$$

where $X$ and $Y$ are the noisy and restored images, respectively. $\partial_{xy}$ is the covariance of $X$ and $Y$ $c_1 = (k_1 L)^2$, $c_2 = (k_2 L)^2$ are two variables to stabilize the division with weak denominator. $L$ is the dynamic range of pixel values (typically this is $(2^{(\#bitsperpixel)} - 1)$). $k_1 = 0.01$ and $k_2 = 0.03$ by default. The measurements or predictions of image quality are based on comparing the restored image with the original, uncorrupted and free of noise image as the reference.

All algorithms are tested on four 8-bit standard gray-scale images: "Boat", "Peppers", "Goldhill" and "Barbara" with size (512×512). PSNR, SSIM and execution time on the named images are reported in Table 1 ~12 based on averaged values over five runs. Some of the images corrupted with salt & pepper noise ranging from 10% to 95% and their restored versions are shown in figures 1~ 18. We tested our method and some state-of-the-art algorithms on images corrupted by salt and pepper noise with density 10% through 95%. For a more meaningful visual comparison, we report only the output of methods including AIFATM and EWA which have close performance to our method. We also report running time of these algorithms. Experimental results show that the proposed method has higher PSNR and SSIM in most of the cases especially in high level noise density. In addition, EWA has the best performance (running time) in comparison to other methods. In offline data processing, such as storing step in image acquisition process in which user's main concentration is to restore corrupted image with high quality as well as preserving visual property, our method outweighs EWA. However, EWA might be used in online applications where quality and preserving visual property does not matter significantly. However, in offline applications our method produces images with better quality.

In order to show how much improvement is obtained by each step of the proposed algorithm, we have presented the PSNR/SSIM values of the images resulting from the first and second step indicated by proposed method-without smoothing (PM-WS) and proposed method (PM) respectively in the results tables. We have also provided a visual example of the first step output in Fig.2.

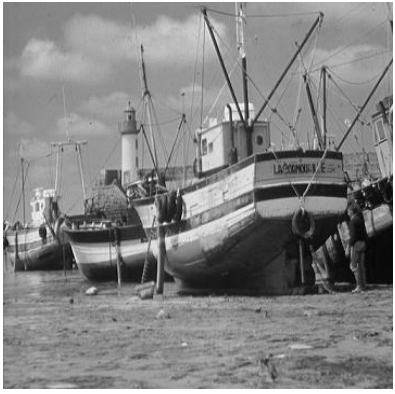 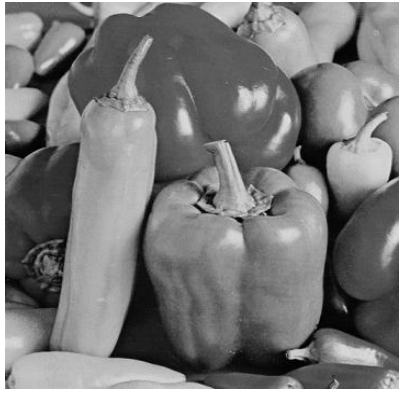 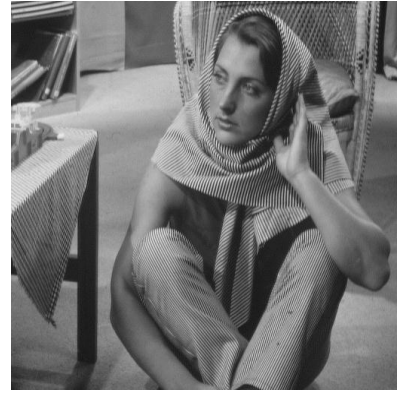

a) Boat original image  b) Peppers original image  c) Barbara original image

**Fig. 1.** Three original images, a) boat, b) peppers and c) barbara used in our experiments.

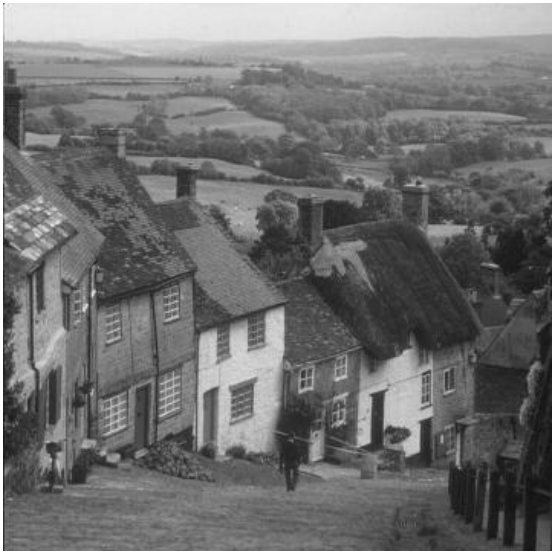 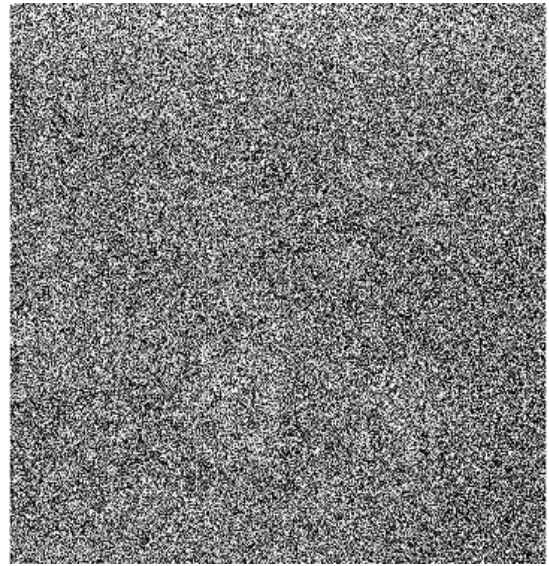

a) Original image  b) Image corrupted by 80%

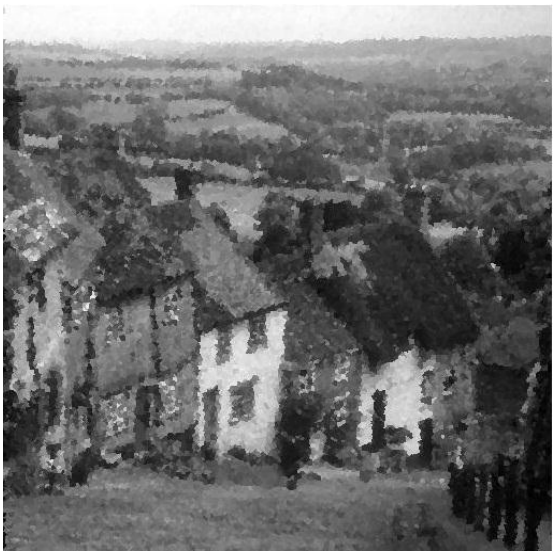 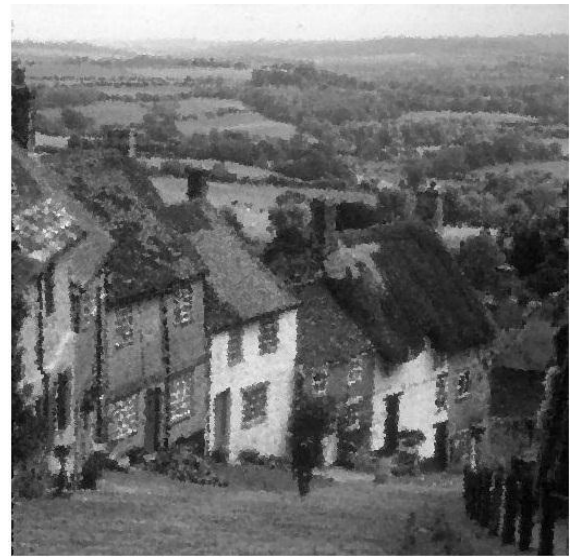

c) Restored image corrupted by 80% (First step)  d) Restored image corrupted by 80% (Second step)

**Fig 2.** Visual comparison between first and second steps over goldhill image corrupted by 80% impulsive noise.

Table 1 Comparing different algorithms by PSNR (db) for "goldhill" from 10% to 95% noise density level. The bold data indicate the method with best performance for the corresponding noise density.

| Noise Density | 10% | 20% | 30% | 40% | 50% | 60% | 70% | 80% | 90% | 95% |
|---|---|---|---|---|---|---|---|---|---|---|
| EDBA | 37.01 | 34.45 | 32.50 | 30.60 | 28.84 | 27.06 | 25.29 | 23.09 | 20.32 | 18.02 |
| NAFSM | 36.68 | 33.55 | 31.76 | 30.37 | 29.51 | 28.50 | 27.52 | 26.23 | 23.10 | 16.85 |
| MDBUTMF | 39.87 | 36.46 | 34.23 | 32.15 | 29.78 | 26.71 | 23.16 | 19.24 | 15.30 | 13.42 |
| EWA | 39.04 | 36.01 | 34.74 | 32.53 | 31.26 | 30.12 | 28.78 | 27.27 | 25.42 | 23.62 |
| AIFATM | 40.45 | 37.33 | 35.43 | 33.62 | 32.09 | 30.64 | 29.15 | 27.42 | 25.48 | 23.66 |
| PM-WS | 38.86 | 36.13 | 34.59 | 32.71 | 30.94 | 29.36 | 28.98 | 26.80 | 25.32 | 23.86 |
| PM | **40.79** | **37.48** | **35.50** | **33.86** | **32.40** | **31.23** | **30.14** | **27.92** | **26.46** | **24.95** |

Table 2 Comparing different algorithms in measure of SSIM for "goldhill" from 10% to 95% noise density.

| Noise Density | 10% | 20% | 30% | 40% | 50% | 60% | 70% | 80% | 90% | 95% |
|---|---|---|---|---|---|---|---|---|---|---|
| EDBA | 0.995 | 0.987 | 0.970 | 0.958 | 0.930 | 0.888 | 0.824 | 0.710 | 0.523 | 0.370 |
| NAFSM | 0.992 | 0.981 | 0.969 | 0.953 | 0.933 | 0.900 | 0.877 | 0.829 | 0.709 | 0.410 |
| MDBUTMF | 0.995 | 0.988 | 0.978 | 0.963 | 0.938 | 0.883 | 0.768 | 0.560 | 0.292 | 0.165 |
| EWA | 0.995 | 0.990 | 0.984 | 0.976 | 0.966 | 0.951 | 0.929 | **0.891** | **0.816** | 0.720 |
| AIFATM | **0.996** | 0.991 | 0.986 | 0.977 | 0.966 | 0.951 | 0.928 | 0.885 | 0.802 | 0.694 |
| PM-WS | 0.993 | 0.990 | 0.983 | 0.976 | 0.962 | 0.948 | 0.924 | 0.845 | 0.799 | 0.698 |
| PM | 0.995 | **0.993** | **0.988** | **0.979** | **0.969** | **0.958** | **0.937** | 0.889 | 0.812 | **0.770** |

Table 3 Comparing different algorithms in execution time (second) for "goldhill" from 10% to 95% noise density.

| Noise Density | 10% | 20% | 30% | 40% | 50% | 60% | 70% | 80% | 90% | 95% |
|---|---|---|---|---|---|---|---|---|---|---|
| EDBA | 1.51 | 1.51 | 1.71 | 1.12 | 1.23 | 1.44 | 1.33 | 1.37 | 1.27 | 1.33 |
| NAFSM | 4.963 | 7.84 | 10.58 | 13.49 | 15.87 | 18.39 | 21.24 | 23.76 | 27.21 | 27.58 |
| MDBUTMF | 12.23 | 21.0 | 29.72 | 37.05 | 44.64 | 51.30 | 55.76 | 59.85 | 52.90 | 43.5 |
| EWA | **0.61** | **0.61** | **0.60** | **0.60** | **0.80** | **0.65** | **0.6** | **0.61** | **0.69** | **0.71** |
| AIFATM | 5.36 | 10.6 | 16.39 | 23.28 | 30.11 | 37.32 | 49.68 | 73.46 | 123.69 | 188.47 |
| PM-WS | 5.76 | 10.8 | 12.70 | 21.13 | 25.99 | 32.79 | 34.10 | 38.28 | 47.40 | 53.64 |
| PM | 7.32 | 13.4 | 18.96 | 23.64 | 27.92 | 34.01 | 36.75 | 40.9 | 49.32 | 56.87 |

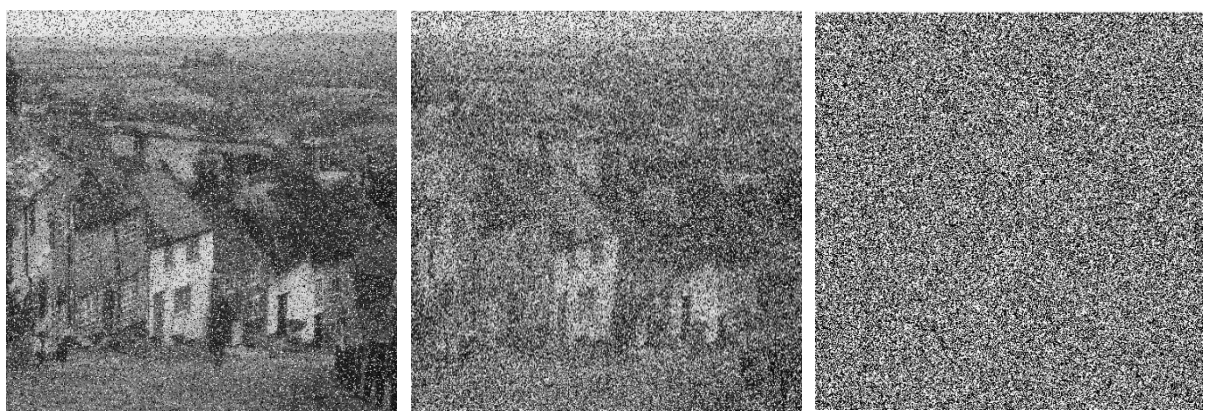

a) Image corrupted by 20%    b) Image corrupted by 50%    c) Image corrupted by 95%

**Fig. 3.** (a, b, c) are goldhill images corrupted by 20%, 50% and 95% impulse noise respectively.

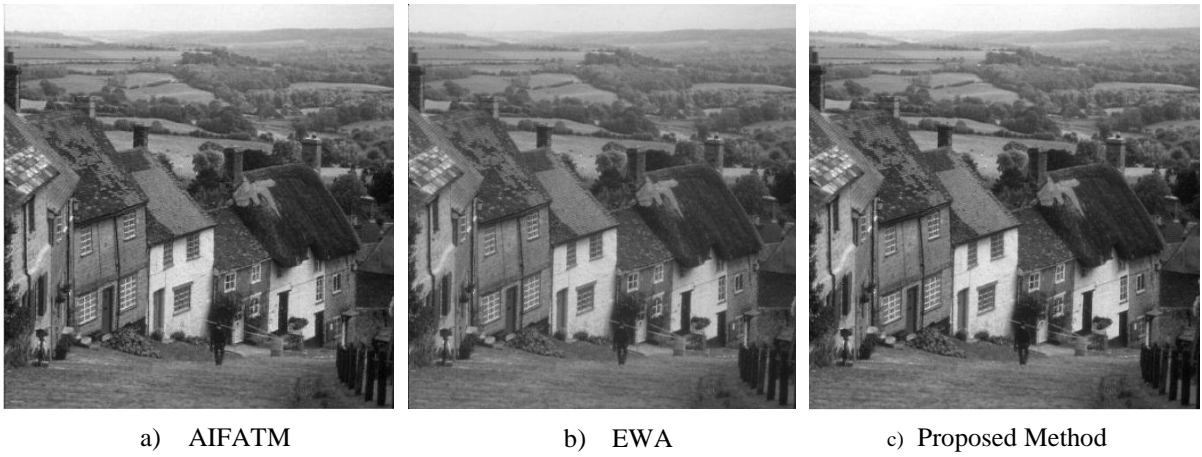

a) AIFATM    b) EWA    c) Proposed Method

**Fig. 4.** The results for goldhill, (a, b, c) show restored images corrupted by 20% using AIFATM, EWA and proposed method.

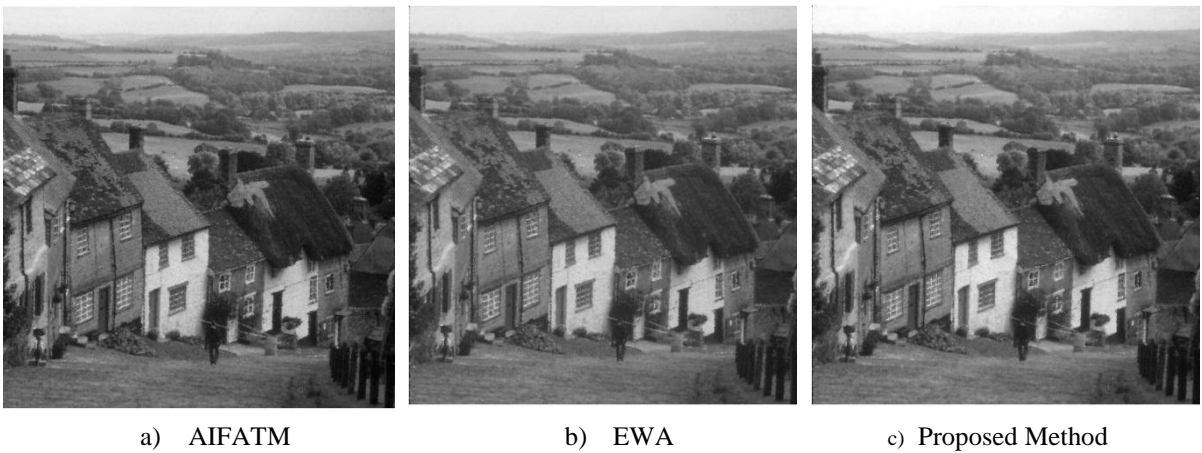

a) AIFATM    b) EWA    c) Proposed Method

**Fig. 5.** The results for goldhill, (a, b, c) show restored images corrupted by 50% using AIFATM, EWA and proposed method.

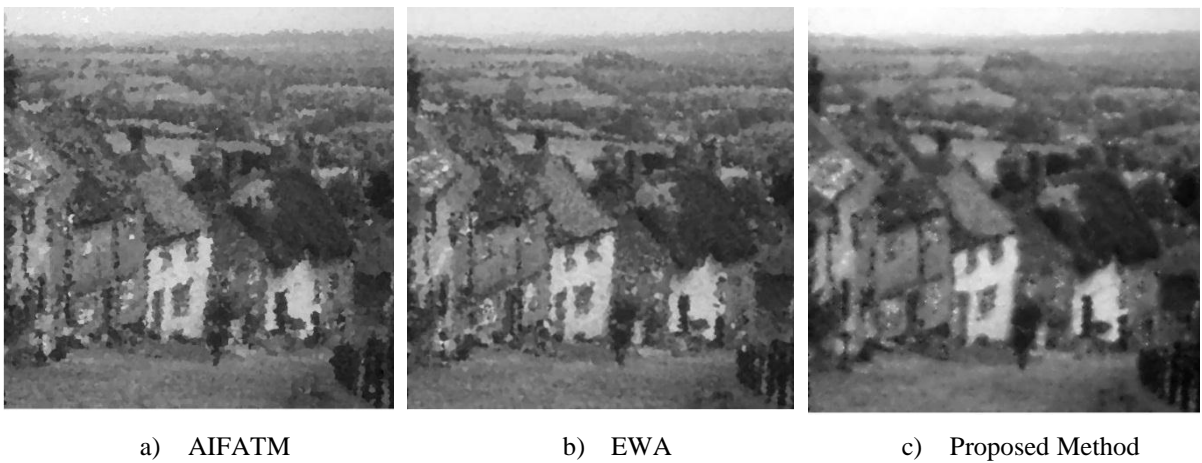

a) AIFATM    b) EWA    c) Proposed Method

**Fig. 6.** The results for goldhill, (a, b, c) show restored images corrupted by 95% using AIFATM, EWA and proposed method.

**Table 4** Comparing different algorithms in measure of PSNR for "boat" from 10% to 95% noise density level.

| Noise Density | 10% | 20% | 30% | 40% | 50% | 60% | 70% | 80% | 90% | 95% |
|---|---|---|---|---|---|---|---|---|---|---|
| EDBA | 38.24 | 34.65 | 32.03 | 29.78 | 27.96 | 25.84 | 23.88 | 21.71 | 18.39 | 16.20 |
| NAFSM | 35.22 | 32.07 | 30.25 | 28.89 | 28.21 | 27.29 | 26.09 | 24.91 | 22.15 | 16.44 |
| MDBUTMF | 39.49 | 36.14 | 33.69 | 31.86 | 29.51 | 26.55 | 22.92 | 19.11 | 15.18 | 13.29 |
| EWA | 38.83 | 35.68 | 33.54 | 32.07 | 30.14 | **29.75** | 28.09 | 26.66 | 23.9 | 22.02 |
| AIFATM | 39.17 | 36.02 | 33.93 | 32.14 | 30.52 | 28.92 | 27.43 | 25.72 | 23.66 | 21.89 |

| | | | | | | | | | |
|---|---|---|---|---|---|---|---|---|---|
| PM-WS | 38.55 | 35.80 | 33.30 | 31.01 | 29.72 | 28.15 | 27.55 | 25.43 | 24.10 | 22.40 |
| **PM** | **39.82** | **36.43** | **34.10** | **32.30** | **30.90** | 29.56 | **28.24** | **26.91** | **25.05** | **23.35** |

**Table 5** Comparing different algorithms in measure of SSIM for "boat" from 10% to 95% noise density level.

| Noise Density | 10% | 20% | 30% | 40% | 50% | 60% | 70% | 80% | 90% | 95% |
|---|---|---|---|---|---|---|---|---|---|---|
| **EDBA** | **0.996** | 0.990 | 0.980 | 0.966 | 0.945 | 0.908 | 0.852 | 0.761 | 0.584 | 0.457 |
| **NAFSM** | 0.993 | 0.984 | 0.973 | 0.960 | 0.941 | 0.920 | 0.887 | 0.846 | 0.726 | 0.440 |
| **MDBUTMF** | **0.996** | **0.991** | 0.983 | 0.970 | 0.941 | 0.883 | 0.743 | 0.532 | 0.285 | 0.160 |
| **EWA** | **0.996** | **0.991** | 0.986 | 0.980 | 0.973 | **0.962** | 0.948 | 0.921 | 0.872 | 0.823 |
| **AIFATM** | 0.995 | 0.990 | 0.984 | 0.975 | 0.963 | 0.947 | 0.923 | 0.880 | 0.795 | 0.689 |
| **PM-WS** | 0.989 | 0.988 | 0.981 | 0.972 | 0.953 | 0.941 | 0.933 | 0.887 | 0.874 | 0.836 |
| **PM** | 0.995 | **0.991** | **0.988** | **0.982** | **0.976** | 0.958 | **0.952** | **0.936** | **0.901** | **0.848** |

**Table 6** Comparing different algorithms in execution time (seconds) for "boat" from 10% to 95% noise density level.

| Noise Density | 10% | 20% | 30% | 40% | 50% | 60% | 70% | 80% | 90% | 95% |
|---|---|---|---|---|---|---|---|---|---|---|
| **EDBA** | 0.94 | 0.95 | 0.98 | 0.98 | 0.97 | 1.01 | 0.98 | 0.96 | 0.99 | 1.01 |
| **NAFSM** | 4.87 | 8.00 | 11.18 | 13.74 | 16.63 | 19.03 | 22.98 | 24.46 | 27.11 | 29.33 |
| **MDBUTMF** | 12.43 | 20.10 | 27.94 | 35.16 | 43.72 | 49.34 | 55.67 | 55.83 | 53.00 | 42.10 |
| **EWA** | **0.55** | **0.62** | **0.66** | **0.611** | **0.60** | **0.612** | **0.69** | **0.68** | **0.70** | **0.67** |
| **AIFATM** | 5.26 | 10.79 | 16.73 | 23.56 | 29.56 | 37.24 | 47.07 | 69.06 | 119.00 | 189.68 |
| **PM-WS** | 4.46 | 10.60 | 16.10 | 20.16 | 25.06 | 30.31 | 34.63 | 40.09 | 48.30 | 54.90 |
| **PM** | 8.26 | 13.72 | 18.99 | 23.61 | 28.56 | 34.31 | 36.68 | 42.21 | 50.10 | 57.40 |

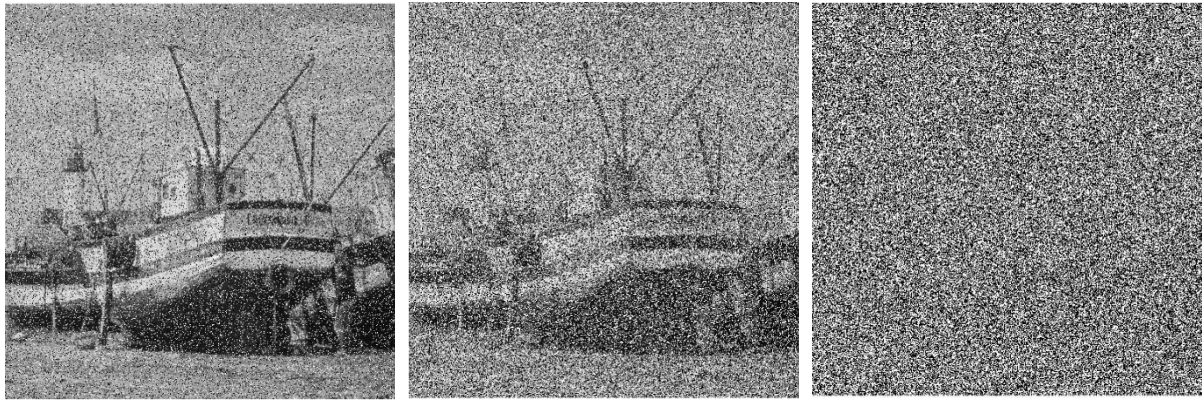

a) Image corrupted by 20%    b) Image corrupted by 50%    c) Image corrupted by 95%

**Fig. 7.** (a, b, c) are boat images corrupted by 20%, 50% and 95% impulse noise respectively.

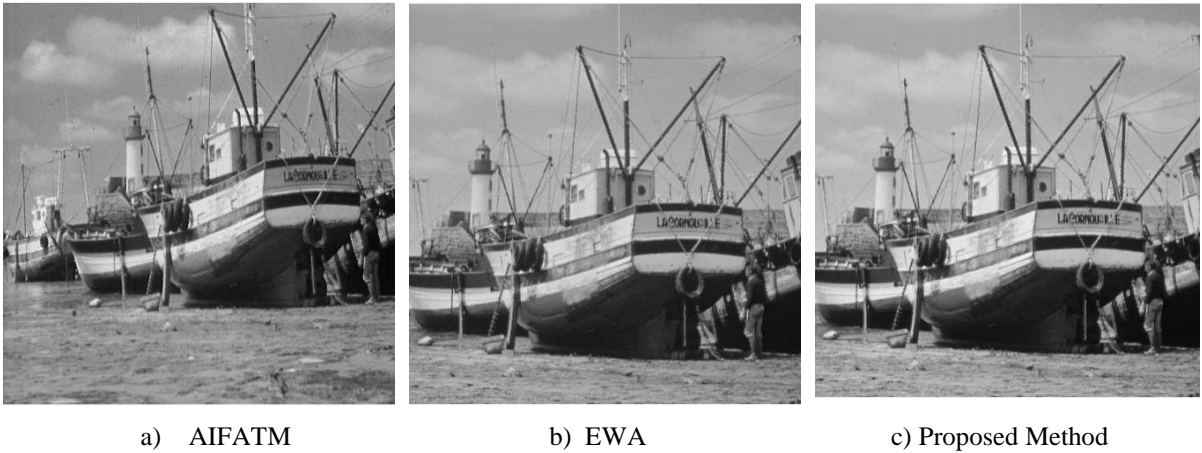

a) AIFATM  b) EWA  c) Proposed Method

**Fig. 8.** The results for boat, (a, b, c) show restored images corrupted by 20% using AIFATM, EWA and proposed method.

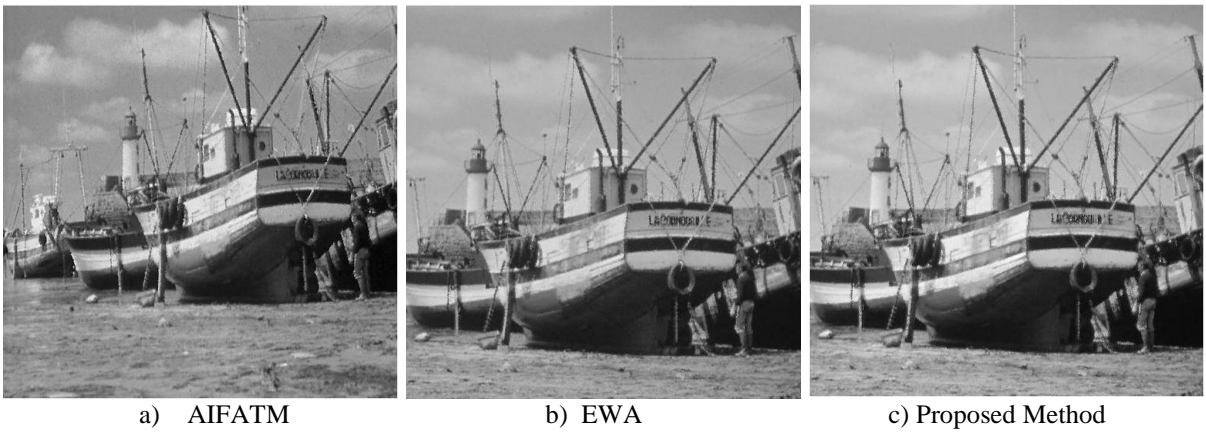

a) AIFATM  b) EWA  c) Proposed Method

**Fig. 9.** The results for boat, (a, b, c) show restored images corrupted by 50% using AIFATM, EWA and proposed method.

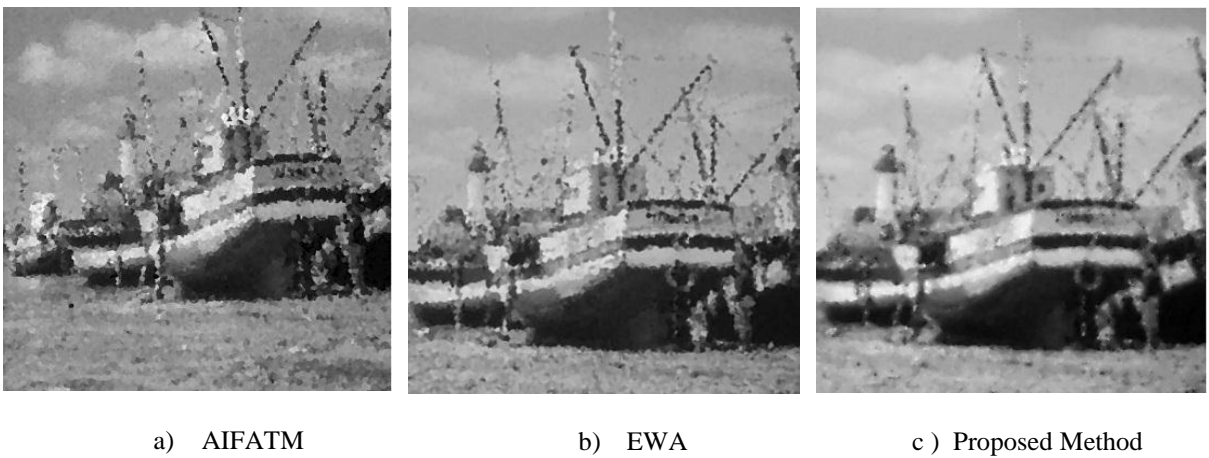

a) AIFATM  b) EWA  c ) Proposed Method

**Fig. 10.** The results for boat, (a, b, c) show restored images corrupted by 95% using AIFATM, EWA and proposed method.

**Table 7** Comparing different algorithms in measure of PSNR (db) for "peppers" from 10% to 95% noise density level.

| Noise Density | 10% | 20% | 30% | 40% | 50% | 60% | 70% | 80% | 90% | 95% |
|---|---|---|---|---|---|---|---|---|---|---|
| EDBA | 39.90 | 36.50 | 34.10 | 31.79 | 29.83 | 27.70 | 25.39 | 22.39 | 18.70 | 16.11 |

| | | | | | | | | | |
|---|---|---|---|---|---|---|---|---|---|
| NAFSM | 38.25 | 35.36 | 33.32 | 32.89 | 31.63 | 30.34 | 29.17 | 27.29 | 23.49 | 16.72 |
| MDBUTMF | 40.85 | 37.33 | 35.05 | 33.14 | 30.77 | 27.35 | 23.30 | 19.211 | 15.18 | 13.14 |
| EWA | 40.39 | 37.29 | 35.34 | 33.74 | 32.52 | 31.12 | 29.71 | 28.26 | 25.93 | 23.82 |
| AIFATM | 41.80 | 38.40 | 36.36 | 34.58 | 33.14 | 31.66 | 30.09 | 28.46 | 26.33 | 24.08 |
| PM-WS | 40.11 | 37.84 | 35.43 | 33.44 | 32.84 | 31.26 | 29.58 | 28.57 | 26.94 | 24.26 |
| **PM** | **41.86** | **38.47** | **36.41** | **34.68** | **33.40** | **32.10** | **30.92** | **29.70** | **27.67** | **25.45** |

**Table 8** Comparing different algorithms in measure of SSIM for "peppers" from 10% to 95% noise density level.

| Noise Density | 10% | 20% | 30% | 40% | 50% | 60% | 70% | 80% | 90% | 95% |
|---|---|---|---|---|---|---|---|---|---|---|
| EDBA | 0.994 | 0.988 | 0.979 | 0.967 | 0.949 | 0.922 | 0.877 | 0.792 | 0.627 | 0.466 |
| NAFSM | 0.994 | 0.988 | 0.981 | 0.972 | 0.963 | 0.950 | 0.933 | 0.900 | 0.787 | 0.450 |
| MDBUTMF | 0.995 | 0.989 | 0.981 | 0.967 | 0.939 | 0.881 | 0.741 | 0.519 | 0.270 | 0.149 |
| EWA | **0.996** | **0.991** | **0.986** | 0.980 | 0.973 | 0.962 | 0.948 | 0.9212 | 0.872 | 0.823 |
| AIFATM | 0.995 | **0.991** | 0.985 | 0.979 | 0.971 | 0.959 | 0.942 | 0.916 | 0.865 | 0.801 |
| PM-WS | 0.994 | 0.98 | 0.978 | 0.963 | 0.954 | 0.950 | 0.929 | 0.919 | 0.890 | 0.829 |
| **PM** | 0.995 | 0.990 | **0.986** | **0.982** | **0.975** | **0.966** | **0.952** | **0.936** | **0.901** | **0.848** |

**Table 9** Comparing different algorithms in execution time (second) for " peppers" from 10% to 95% noise density level.

| Noise Density | 10% | 20% | 30% | 40% | 50% | 60% | 70% | 80% | 90% | 95% |
|---|---|---|---|---|---|---|---|---|---|---|
| EDBA | 0.95 | 1.02 | 1.02 | 0.98 | 1.01 | 0.95 | 1.04 | 0.99 | 1.05 | 0.99 |
| NAFSM | 5.02 | 7.44 | 10.19 | 12.95 | 15.70 | 18.74 | 21.39 | 24.04 | 28.15 | 28.83 |
| MDBUTMF | 11.77 | 19.99 | 28.00 | 34.98 | 42.37 | 48.39 | 53.56 | 54.72 | 49.08 | 41.03 |
| EWA | **0.55** | **0.62** | **0.66** | **0.611** | **0.60** | **0.612** | **0.69** | **0.68** | **0.70** | **0.67** |
| AIFATM | 4.72 | 10.14 | 15.79 | 21.96 | 29.65 | 36.46 | 47.77 | 70.89 | 119.90 | 174.49 |
| PM-WS | 4.16 | 10.38 | 16.60 | 21.26 | 26.87 | 30.64 | 32.69 | 38.33 | 45.81 | 55.83 |
| **PM** | 7.82 | 13.52 | 18.52 | 23.57 | 28.06 | 32.29 | 35.85 | 40.62 | 47.47 | 57.08 |

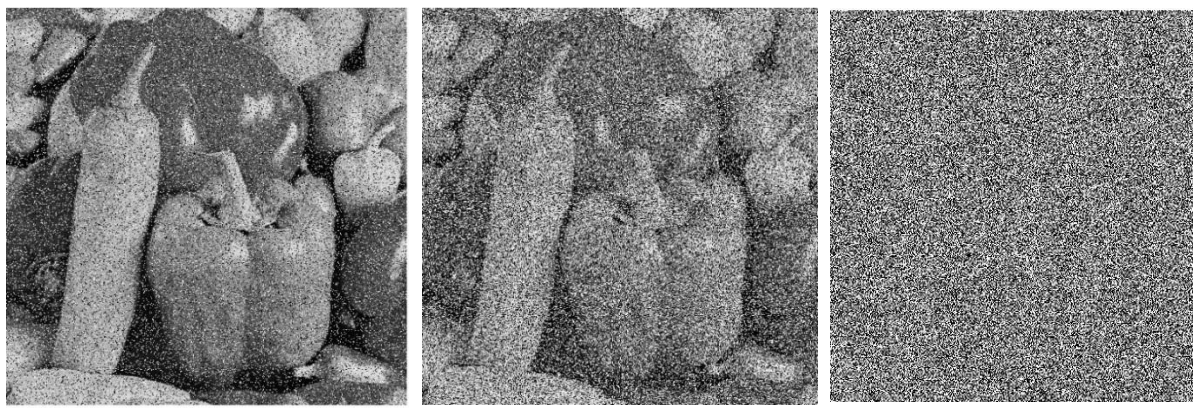

a) Image corrupted by 20%     b) Image corrupted by 50%     c) Image corrupted by 95%

**Fig. 11.** (a, b, c) are peppers images corrupted by 20%, 50% and 95% impulse noise respectively.

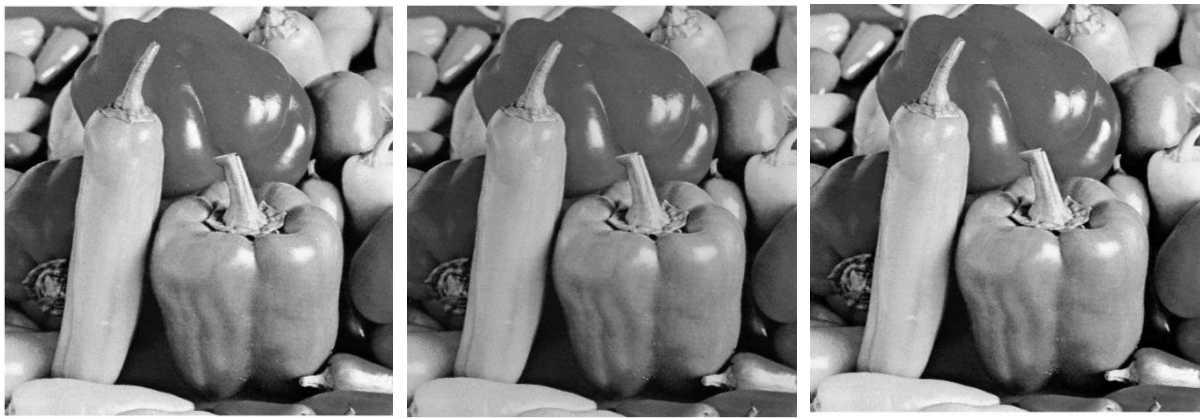

a) AIFATM  b) EWA  c) Proposed Method

**Fig. 12.** The results for peppers, (a, b, c) show restored images corrupted by 20% using AIFATM, EWA and proposed method.

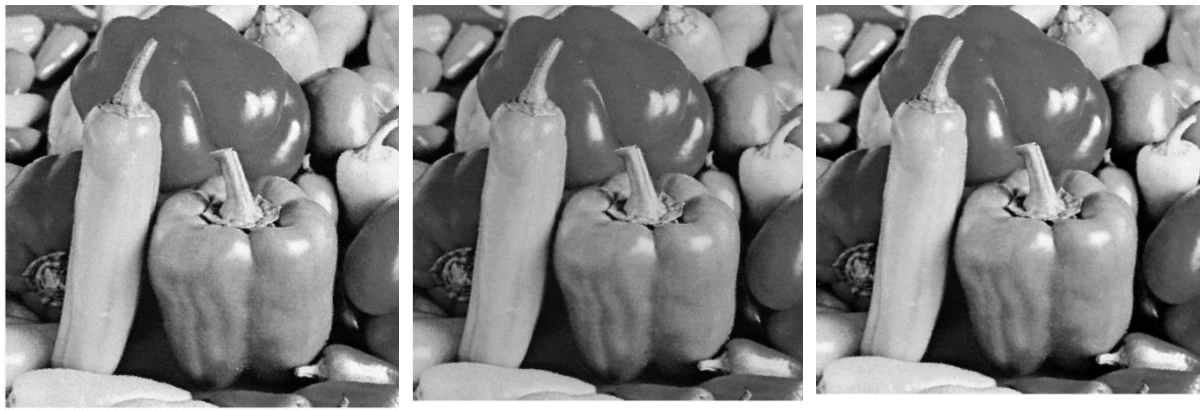

a) AIFATM  b) EWA  c) Proposed Method

**Fig. 13.** The results for peppers, (a, b, c) show restored images corrupted by 50% using AIFATM, EWA and proposed method.

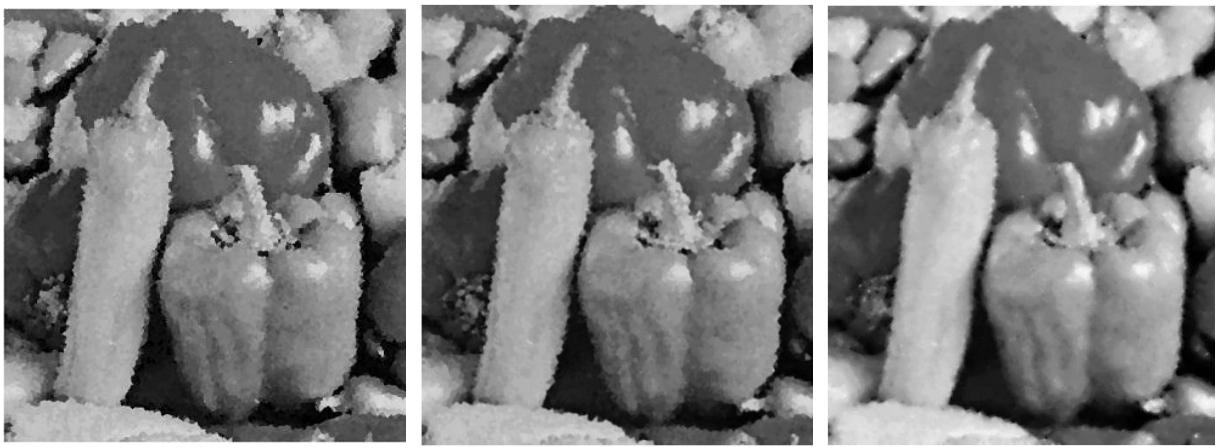

a) AIFATM  b) EWA  c) Proposed Method

**Fig. 14.** The results for peppers, (a, b, c) show restored images corrupted by 95% using AIFATM, EWA and proposed method.

**Table 10** Comparing different algorithms in measure of PSNR (db) for "Barbara" from 10% to 95% noise density level.

| Noise Density | 10% | 20% | 30% | 40% | 50% | 60% | 70% | 80% | 90% | 95% |
|---|---|---|---|---|---|---|---|---|---|---|
| **EDBA** | 32.96 | 29.76 | 27.65 | 25.96 | 24.54 | 23.21 | 21.74 | 19.93 | 17.49 | 15.24 |
| **NAFSM** | 33.15 | 30.07 | 28.23 | 26.88 | 25.75 | 24.74 | 23.82 | 22.73 | 20.74 | 15.77 |
| **MDBUTMF** | 33.58 | 30.44 | 28.37 | 26.81 | 25.49 | 23.83 | 21.55 | 18.35 | 14.78 | 12.99 |
| **EWA** | 34.08 | 30.99 | 29.12 | 27.76 | 26.55 | 25.34 | 24.33 | 23.16 | 21.79 | 20.69 |
| **AIFATM** | 34.88 | 31.71 | 29.71 | 28.1 | 26.72 | 25.36 | 24.08 | 22.75 | 21.29 | 20.32 |
| **PM-WS** | 33.86 | 30.48 | 29.93 | 27.93 | 26.16 | 25.56 | 24.75 | 23.30 | 21.92 | 20.94 |
| **PM** | **34.97** | **32.71** | **30.84** | **29.27** | **27.09** | **26.09** | **25.25** | **23.46** | **22.52** | **21.74** |

**Table 11** Comparing different algorithms in measure of SSIM for "Barbara" from 10% to 95% noise density level.

| Noise Density | 10% | 20% | 30% | 40% | 50% | 60% | 70% | 80% | 90% | 95% |
|---|---|---|---|---|---|---|---|---|---|---|
| **EDBA** | 0.987 | 0.970 | 0.949 | 0.922 | 0.890 | 0.849 | 0.782 | 0.684 | 0.502 | 0.345 |
| **NAFSM** | 0.988 | 0.974 | 0.958 | 0.938 | 0.914 | 0.882 | 0.846 | 0.796 | 0.690 | 0.412 |
| **MDBUTMF** | 0.988 | 0.974 | 0.957 | 0.935 | 0.906 | 0.849 | 0.728 | 0.515 | 0.273 | 0.160 |
| **EWA** | 0.989 | 0.977 | 0.963 | **0.942** | 0.926 | 0.898 | 0.869 | 0.826 | 0.756 | 0.683 |
| **AIFATM** | **0.990** | **0.979** | 0.966 | 0.938 | 0.927 | 0.902 | 0.867 | 0.819 | 0.740 | 0.668 |
| **PM-WS** | 0.985 | 0.97 | 0.968 | 0.939 | 0.919 | 0.907 | 0.885 | 0.822 | 0.759 | 0.692 |
| **PM** | 0.988 | 0.976 | **0.970** | **0.942** | **0.931** | **0.918** | **0.908** | **0.831** | **0.776** | **0.719** |

**Table 12** Comparing different algorithms in execution time (second) for "Barbara" from 10% to 95% noise density level.

| Noise Density | 10% | 20% | 30% | 40% | 50% | 60% | 70% | 80% | 90% | 95% |
|---|---|---|---|---|---|---|---|---|---|---|
| **EDBA** | 11.03 | 11.29 | 10.71 | 10.62 | 10.95 | 15.22 | 11.14 | 12.17 | 10.29 | 10.53 |
| **NAFSM** | 3.6 | 6.73 | 9.89 | 13.1 | 17.43 | 20.40 | 23.55 | 26.53 | 28.59 | 29.77 |
| **MDBUTMF** | 5.3 | 6.2 | 7.6 | 9.34 | 10.7 | 11.93 | 13.74 | 14.15 | 13.23 | 12.30 |
| **EWA** | **0.90** | **0.87** | **0.85** | **0.89** | **0.94** | **0.87** | **1.02** | **0.88** | **0.99** | **1.05** |
| **AIFATM** | 2.17 | 4.37 | 6.57 | 8.62 | 11.01 | 13.87 | 18.61 | 26.57 | 49.55 | 87.70 |
| **PM-WS** | 8.14 | 14.11 | 20.52 | 25.45 | 29.14 | 35.05 | 43.00 | 45.52 | 57.60 | 59.40 |
| **PM** | 10.56 | 17.40 | 24.12 | 27.10 | 32.20 | 38.60 | 46.70 | 48.35 | 59.42 | 61.10 |

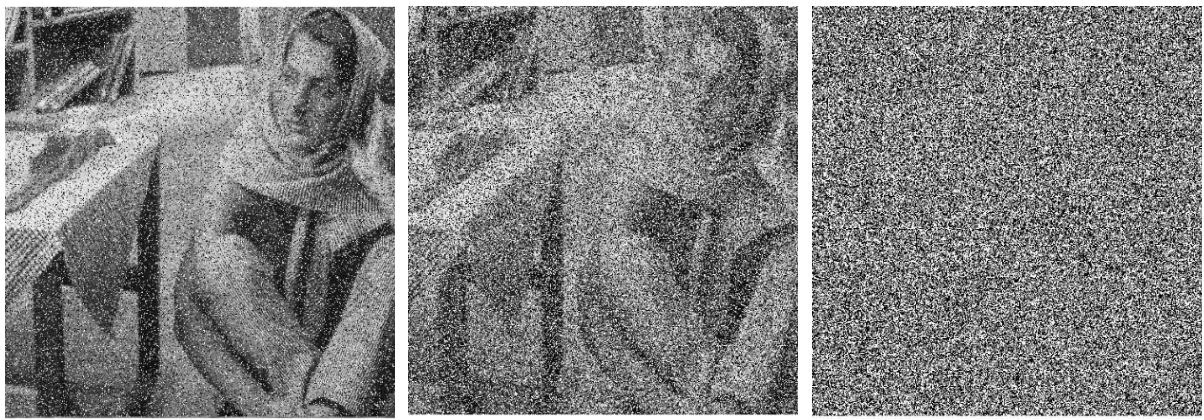

a) Image corrupted by 20%    b) Image corrupted by 50%    c) Image corrupted by 95%

**Fig. 15.** (a, b, c) are peppers images corrupted by 20%, 50% and 95% impulse noise respectively.

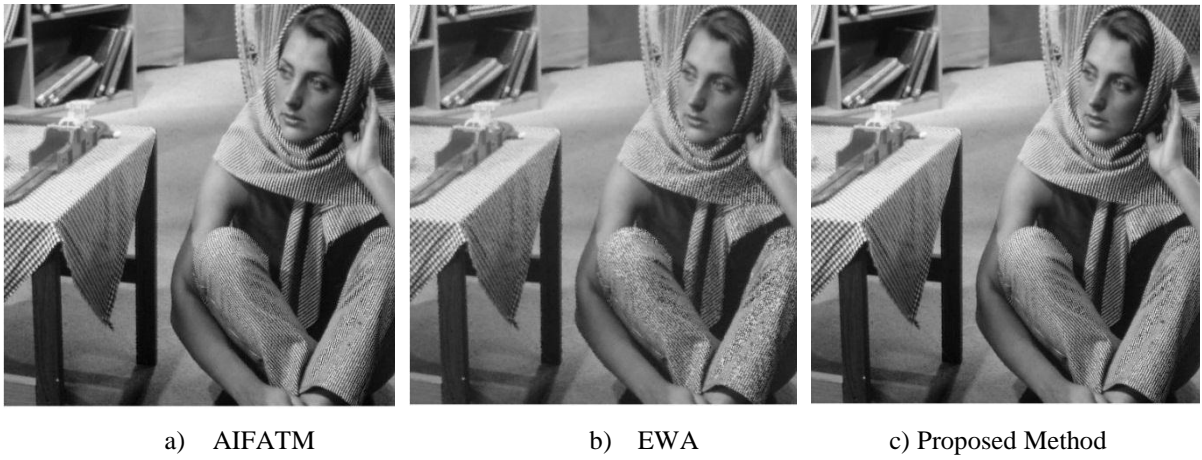

a) AIFATM  b) EWA  c) Proposed Method

**Fig. 16.** The results for peppers, (a, b, c) show restored images corrupted by 20% using AIFATM, EWA and proposed method.

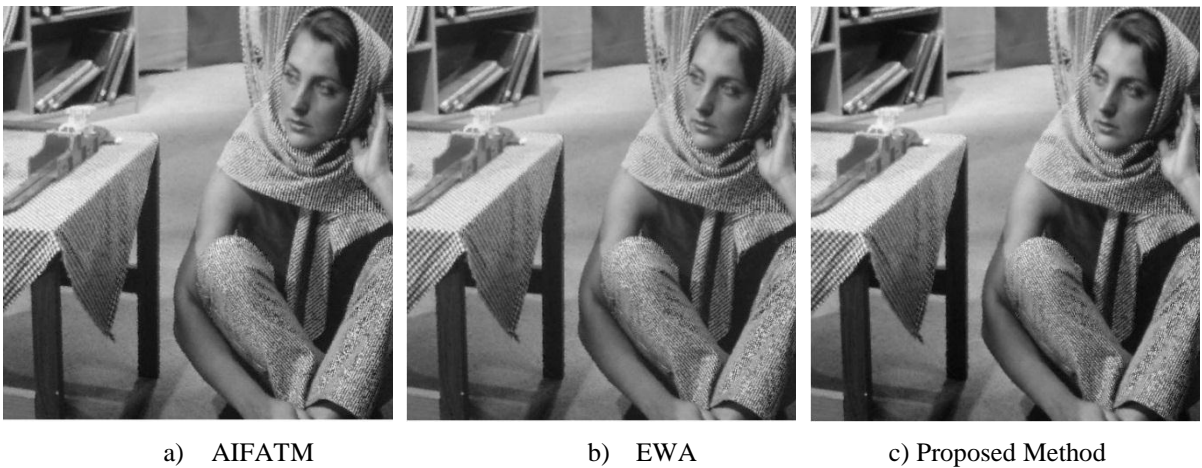

a) AIFATM  b) EWA  c) Proposed Method

**Fig. 17.** The results for peppers, (a, b, c) show restored images corrupted by 50% using AIFATM, EWA and proposed method.

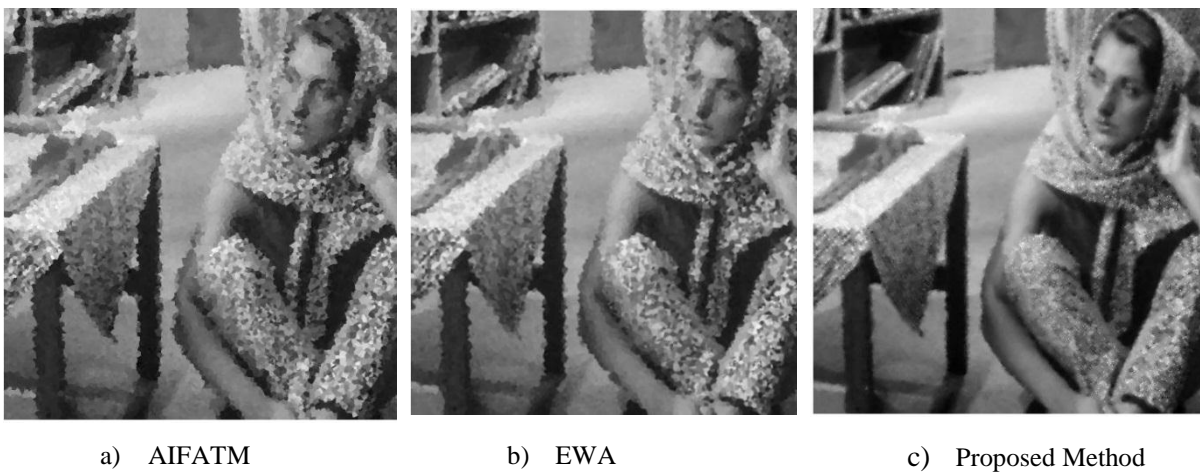

a) AIFATM  b) EWA  c) Proposed Method

**Fig. 18.** The results for peppers, (a, b, c) show restored images corrupted by 95% using AIFATM, EWA and proposed method.

## Conclusion

In this work, we proposed a new algorithm for restoring images highly corrupted by impulse noise. To estimate the proper intensity value of a noisy pixel using RBF, we fit a continuous model on a small patch centered on the noisy pixel. Then we smooth the reconstructed image to overcome the artifacts and residual error in estimation resulted from interpolation step. Our method showed better results in measures of PSNR and SSIM in comparison to recent effective algorithms. Experimental results show that superiority of the proposed method is in the high noise ratio. Execution time of the proposed method shows this method is suitable for offline applications such as storing step in the image acquisition process which restores images with better quality and preserves details such as thin lines in comparison to some recent effective methods. In addition to higher average PSNR and SSIM, our method restores images with higher visual quality and provides better edge and texture preservation compared to state of the art algorithms. This method also does not require complex impulse detector. Another advantage of the proposed method is that for a given noisy image, there is no need to tune parameters by trial and error to achieve the best result. In future, we look forward to extend our method to apply on different type of noises such as Gaussian and Speckle.